\documentclass[twocolumn,pre,superscriptaddress]{revtex4}
\usepackage{graphics}
\usepackage{graphicx}
\usepackage{amsfonts}
\usepackage{textcomp}
\usepackage{amssymb}
\usepackage{mathrsfs}
\usepackage{amsmath}
\usepackage{color}
\usepackage{float}
\usepackage{soul}

\begin{document}

\title{Active Hyperuniform Networks of Chiral Magnetic Micro-Robotic Spinners}

\author{Jing Wang\footnote{These authors contributed equally to this work.}}
\affiliation{Wenzhou Institute, University of Chinese Academy of Sciences, Wenzhou 325001 Zhejiang, China}

\author{Zihao Sun\footnotemark[1]}
\affiliation{School of Physical Sciences, University of Chinese Academy of Sciences, Beijing 100049, China}

\author{Huaicheng Chen}
\affiliation{Wenzhou Institute, University of Chinese Academy of Sciences, Wenzhou 325001 Zhejiang, China}

\author{Gao Wang}
\affiliation{Wenzhou Institute, University of Chinese Academy of Sciences, Wenzhou 325001 Zhejiang, China}

\author{Duyu Chen}
\affiliation{Materials Research Laboratory, University of California, Santa Barbara 93106 California, USA}

\author{Guo Chen}
\affiliation{Wenzhou Institute, University of Chinese Academy of Sciences, Wenzhou 325001 Zhejiang, China}
\affiliation{Chongqing Key Laboratory of Interface Physics in Energy Conversion, College of Physics, Chongqing University, Chongqing 401331, China}

\author{Jianwei Shuai}
\affiliation{Wenzhou Institute, University of Chinese Academy of Sciences, Wenzhou 325001 Zhejiang, China}

\author{Mingcheng Yang}
\email[correspondence sent to: ]{mcyang@iphy.ac.cn}
\affiliation{School of Physical Sciences, University of Chinese Academy of Sciences, Beijing 100049, China}

\author{Yang Jiao}
\email[correspondence sent to: ]{yang.jiao.2@asu.edu}
\affiliation{Materials Science and Engineering, Arizona State University, Tempe 85287 AZ, USA} \affiliation{Department of Physics, Arizona State University, Tempe 85287 AZ, USA}

\author{Liyu Liu}\email[correspondence sent to: ]{liu@iphy.ac.cn}
\affiliation{Human Phenome Institute, Fudan University, Shanghai 201203, China}
\affiliation{Wenzhou Institute, University of Chinese Academy of Sciences, Wenzhou 325001 Zhejiang, China}


\begin{abstract}
Disorder hyperuniform (DHU) systems possess a hidden long-range order manifested as the complete suppression of normalized large-scale density fluctuations like crystals, which endows them with many unique properties. Here, we demonstrate a new organization mechanism for achieving stable DHU structures in active-particle systems via investigating the self-assembly of robotic spinners with three-fold symmetric magnetic binding sites up to a heretofore experimentally unattained system size, i.e., with $\sim 1000$ robots. The spinners can self-organize into a wide spectrum of actively rotating three-coordinated network structures, among which a set of stable DHU networks robustly emerge. These DHU networks are topological transformations of a honeycomb network by continuously introducing the Stone-Wales defects, which are resulted from the competition between tunable magnetic binding and local twist due to active rotation of the robots. Our results reveal novel mechanisms for emergent DHU states in active systems and achieving novel DHU materials with desirable properties.
\end{abstract}
\maketitle

\section{Introduction}

Disorder hyperuniformity (DHU) is a recently discovered exotic state of many-body systems \cite{To03, To18a}, characterized by an unusual combination of amorphous local structures as in liquids or glasses, and a hidden long-range order manifested as strong suppression of normalized large-scale density fluctuations like a perfect crystal \cite{To03, To18a}. This unique ``glass-crystal'' duality can endow DHU systems with novel physical properties that are traditionally thought to be unattainable \cite{Fl09, Fl13, Zh16, Ch18a, Xu17}. One example is the discovery of novel DHU materials with tunable large, fully isotropic, complete photonic \cite{Fl09} and phononic \cite{Fl13} bandgaps, enabling new wave manipulation applications \cite{klatt2022wave, Le16, yu2023evolving, shi2023computational} and devices \cite{park2024deep, wang2025optimization, diego2025hypersonic}. Thus, engineering and experimental realization of DHU structures serve as a crucial first step in developing DHU materials with desirable properties.


Despite the fact that DHU states have been observed in a wide spectrum of physical \cite{Fe56, Ge19quantum, Ru19, Sa19, sakai2022quantum, sanchez2023disordered, chen2025anomalous, Zh20, chen2023disordered, zhang2023approach, Ch21, chen2021multihyperuniform, Ch18b, Ku11, Hu12, Dr15, Do05, Za11a, yuan2021universality, Ga02, atkinson2012maximally} and biological \cite{Ji14, Ma15, liu2024universal, ge2023hidden} systems across scales, the experimental realization of DHU materials with controllable structural features still poses many challenges. For example, 3D printing techniques \cite{chen20243d} are limited by resolutions and are typically not scalable. Achieving DHU states under equilibrium conditions requires highly complex long-range interactions, which are very difficult to produce experimentally \cite{To15, jiao2022hyperuniformity}. DHU states obtained via non-equilibrium routes \cite{He15, Ja15, We15, salvalaglio2020hyperuniform, nizam2021dynamic, zheng2023universal, ma2023theory} such as random organization \cite{hexner2017noise, hexner2017enhanced, weijs2017mixing, wilken2022random} and active fluidization \cite{Le19, lei2019hydrodynamics, huang2021circular, zhang2022hyperuniform, oppenheimer2022hyperuniformity} typically do not form stable solid structures for material applications. Very recently, DHU state in the absorbing phase of active binary mixtures of non-reciprocally interacting robots (with $N \sim 50$) was experimentally observed \cite{chen2024emergent}.

Here, we demonstrate a new organization mechanism for achieving stable DHU structures in active-particle systems. Specifically, we report robust experimental realization of a variety of rotating (i.e., ``active'') disordered hyperuniform network structures via self-organization of magnetic robotic spinners at a heretofore un-attained system size $N$, i.e., up to $N \sim 1000$ robots. Such a system size allows us to access much smaller wavenumbers compared to previous studies, which is crucial to accurately ascertaining hyperuniformity. Each robot possesses three symmetrically distributed magnetic binding sites with tunable binding force $F$ and can rotate (clock-wisely) with tunable rotation speed $\omega$ driven by a programmable light field (see Fig. \ref{fig_1}) \cite{wang2024robo}. 

Using both experiments and simulations, we show the competition of tunable magnetic binding and rotation-induced local twist gives rise to a wide spectrum of actively rotating three-coordinated network structures, among which a set of stable disordered hyperuniform networks robustly emerge. These stable DHU networks, which can be considered as topological transformations of a perfect honeycomb network by continuously introducing the Stone-Wales (SW) topological defects \cite{Zh20, Ch21}, possess varying large-scale structural characteristics (e.g,  hyperuniformity exponents \cite{To18a}), controlled by the tunable magnetic force $F$ and rotation speed $\omega$. Moreover, we demonstrate a robust ``phase transition'' from nonhyperuniform to hyperuniform state induced by decreasing the rotation speed. 







\section{Results}

\subsection{Definition of hyperuniformity}

DHU systems possess a local number variance $\sigma_N^2(R)$ within a spherical window of radius $R$ that grows more slowly than the window volume ($\sim R^d$ in $d$-dimensional space) in the large-$R$ limit \cite{To03, To18a}, i.e., $\lim_{R\rightarrow \infty} \sigma_N^2(R)/R^d = 0$. This is satisfied if the static structure factor $S({\bf k})$ vanishes in the zero-wavevector limit, i.e., 
$\lim_{|{\bf k}|\rightarrow 0}S({\bf k}) = 0$,
where ${\bf k}$ is the wavevector and $S({\bf k})$ is related to the pair-correlation function $g_2({\bf r})$ via $S({\bf k}) = 1+\rho \int e^{-i{\bf k}\cdot {\bf r}}[g_2({\bf r}) - 1]d{\bf r}$ and $\rho = N/V$ is the number density. For statistically isotropic systems, the structure factor only depends on the wavenumber $k = |{\bf k}|$. The small-$k$ scaling behavior of $S(k)$, i.e., $S({k}) \sim k^\alpha$, where $\alpha$, called the {\it hyperuniformity exponent},  determines the large-$R$ asymptotic behavior of $\sigma_N^2(R) \sim R^\beta$, based on which all DHU
systems can be categorized into three classes:
$\sigma_N^2(R) \sim R^{d-1}$ for $\alpha>1$ (class I); $\sigma_N^2(R)
\sim R^{d-1}\ln(R)$ for $\alpha=1$ (class II); and $\sigma_N^2(R)
\sim R^{d-\alpha}$ for $0<\alpha<1$ (class III) \cite{To18a}. By contrast, a standard \textit{nonhyperuniform} system (such as a typical liquid or ideal gas) possesses a scaling $\sigma_N^2(R)
\sim R^{d}$. 


\vspace{-0.5cm}
\begin{figure}[ht!]
\begin{center}
$\begin{array}{c}\\
\includegraphics[width=0.485\textwidth]{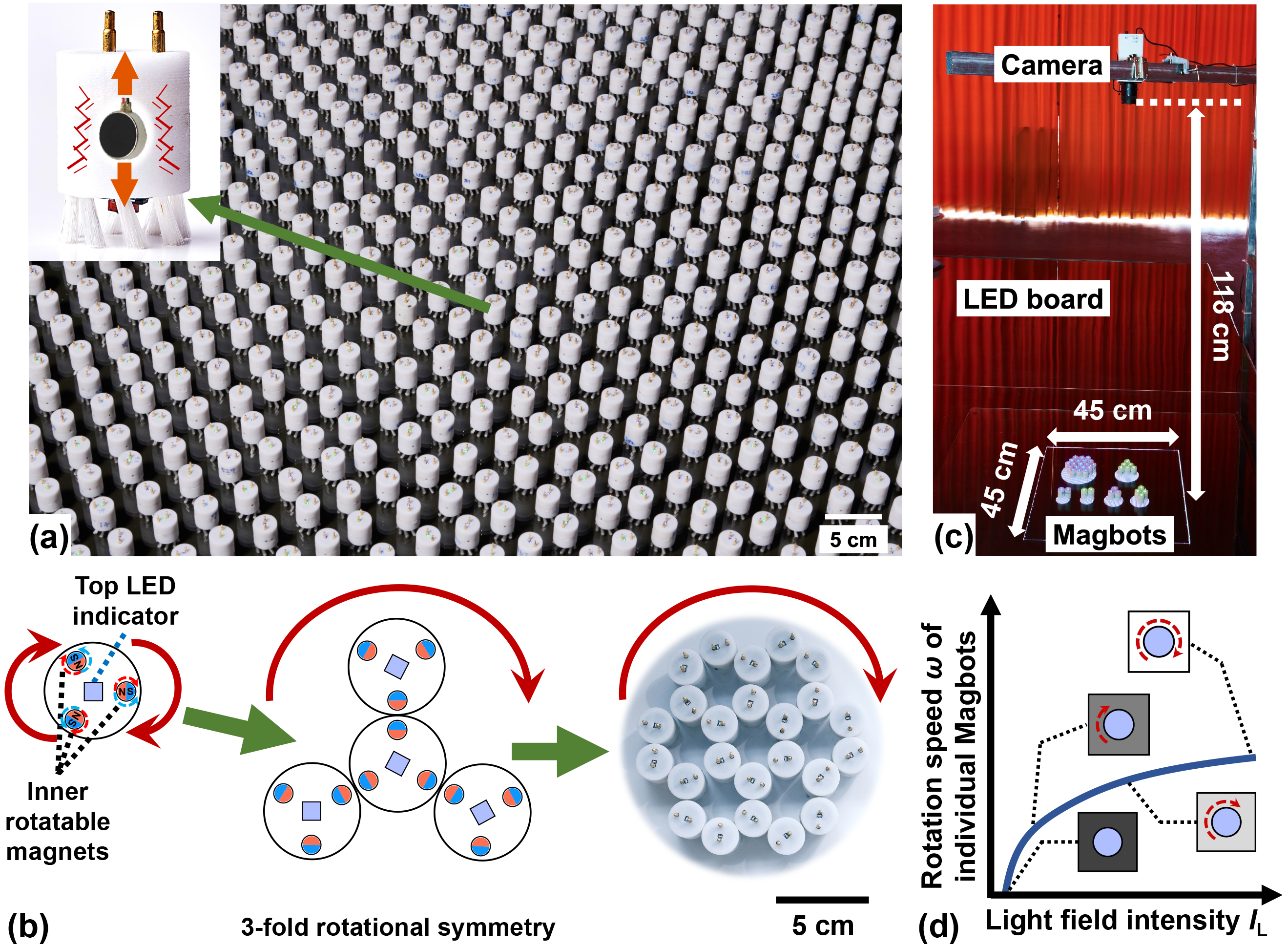} 
\end{array}$
\end{center}
\vspace{-0.7cm}
\caption{(a) Magnetic robot system containing $\sim 1000$ spinners, i.e., the ``Magbots" (inset). (b) 3-fold symmetric magnetic binding sites leading to the assembly of an active cluster of three-coordinated network. (c) An interactive light intensity platform to control the rotation speed of individual Magbots. (d) Rotation speed $\omega$ vs. the intensity $I_L$ of the underlying light field.} \label{fig_1}
\end{figure}


\subsection{Magbots: fundamental building blocks}

The magnetic spinners studied here are coin-sized cylinder-shape robots (inset of Fig. \ref{fig_1}a), which are henceforth referred to as ``Magbots'' \cite{wang2024robo}. A Magbot is composed of a vibration motor whose vertical vibration, in combination of the tilted brushes at the bottom, drives the Magbot to rotate clockwise (CW) (see SI Sec. 1) and generates diffusive translational motion (SI Sec. 2). Each Magbot has 3 ``pockets'', symmetrically distributed on its perimeter, each can host a freely rotating magnetic rod, leading to a locally three-coordinated self-assembly structure under favorable conditions (see Fig. \ref{fig_1}b). The magnetic binding force $F$ is tuned by replacing the magnetic rods.

The Magbots possess light sensors at the bottom, allowing them to respond to a light intensity field (via a LED array, Fig. \ref{fig_1}c). The dynamic states of the Magbots are collected using an overhead low-latency CCD camera. The homogeneous light field plays a similar role of ``temperature'', e.g., increasing light intensity $I_{\rm L}$ leads to faster rotation speeds $\omega$ of the Magbots (see Fig. \ref{fig_1}d and SI Sec. 2), similar to the fact that increasing temperature leads to stronger thermal motions of colloids. As shown below, the tunable magnetic forces (favoring local order) and the twist effect due to active rotation (favoring disordered states) provide two competing effects that give rise to a wide spectrum actively rotating self-organizing network structures.


\vspace{-0.5cm}
\begin{figure}[ht!]
\begin{center}
$\begin{array}{c}\\
\includegraphics[width=0.495\textwidth]{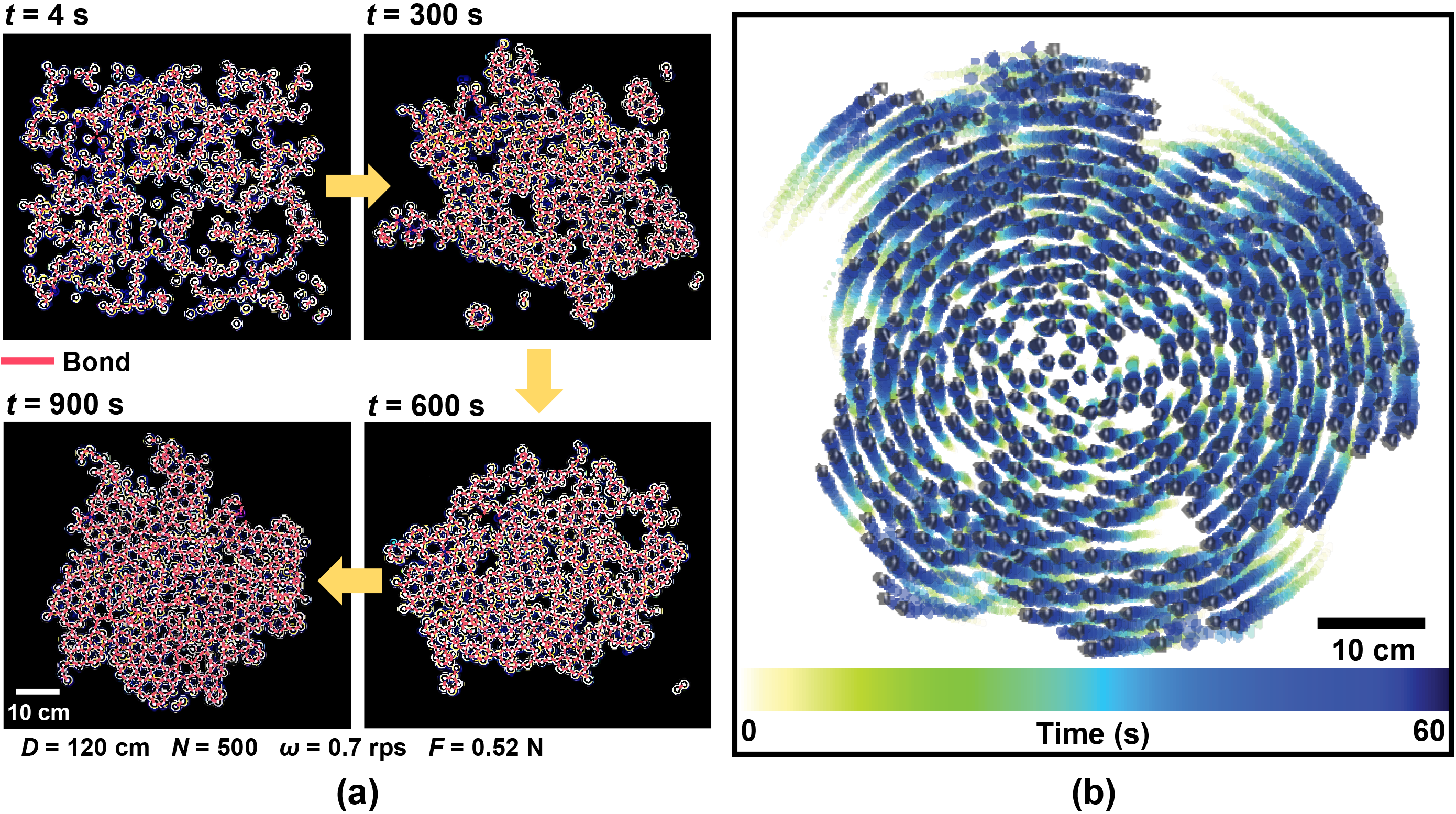} 
\end{array}$
\end{center}
\vspace{-0.7cm}
\caption{(a) Snapshots showing the self-assembly of $N=500$ Magbots with $\omega = 0.7$ rps and $F = 0.52$ N from an initial disordered dispersion state to an active rotating network structure. (b) Illustration of the rotating active assembly structure of the Magbots.} \label{fig_2}
\end{figure}



\subsection{Self-organization of Magbots into actively rotating network structures}

We systematically investigate the self-organization behavior of Magbot systems containing up to $N \sim 1000$ robots, which is significantly larger than previously studied swarm robotic systems \cite{berlinger2021implicit, wang2021emergent, savoie2019robot, li2021programming}. We employ the magnetic binding force $F$ and the rotation speed $\omega$ as control parameters, which respectively play a similar role of enthalpy and kinetic effects as in thermal systems. We select 3 representative values for each parameters, covering the range of experimentally realizable values of $F$ and $\omega$. 



A representative self-assembly process of $N = 500$ Magbots for $F = 0.52$ N and $\omega = 0.7$ rps is shown in Fig. \ref{fig_2}a (also see SI videos). As the Magbots move and collide, they start to form three-coordinated clusters, which percolate, re-arrange, and eventually form a dense, structurally stable three-coordinated network. In the process of self-assembling into a honeycomb lattice \cite{mao2013entropy}, the twist effect due to active rotation of the Magbots frustrate the ordering of the system, preventing the formation of a perfect honeycomb network (see Appendix). The self-organized network is also rotating (Fig. \ref{fig_2}b), which we refer to as an ``active'' network. The experiments are repeated three times for each set of parameters and for varying system size. Robust self-organization of three-coordinated network structures are repeatedly observed.




\vspace{-0.5cm}
\begin{figure}[ht!]
\begin{center}
$\begin{array}{c}\\
\includegraphics[width=0.475\textwidth]{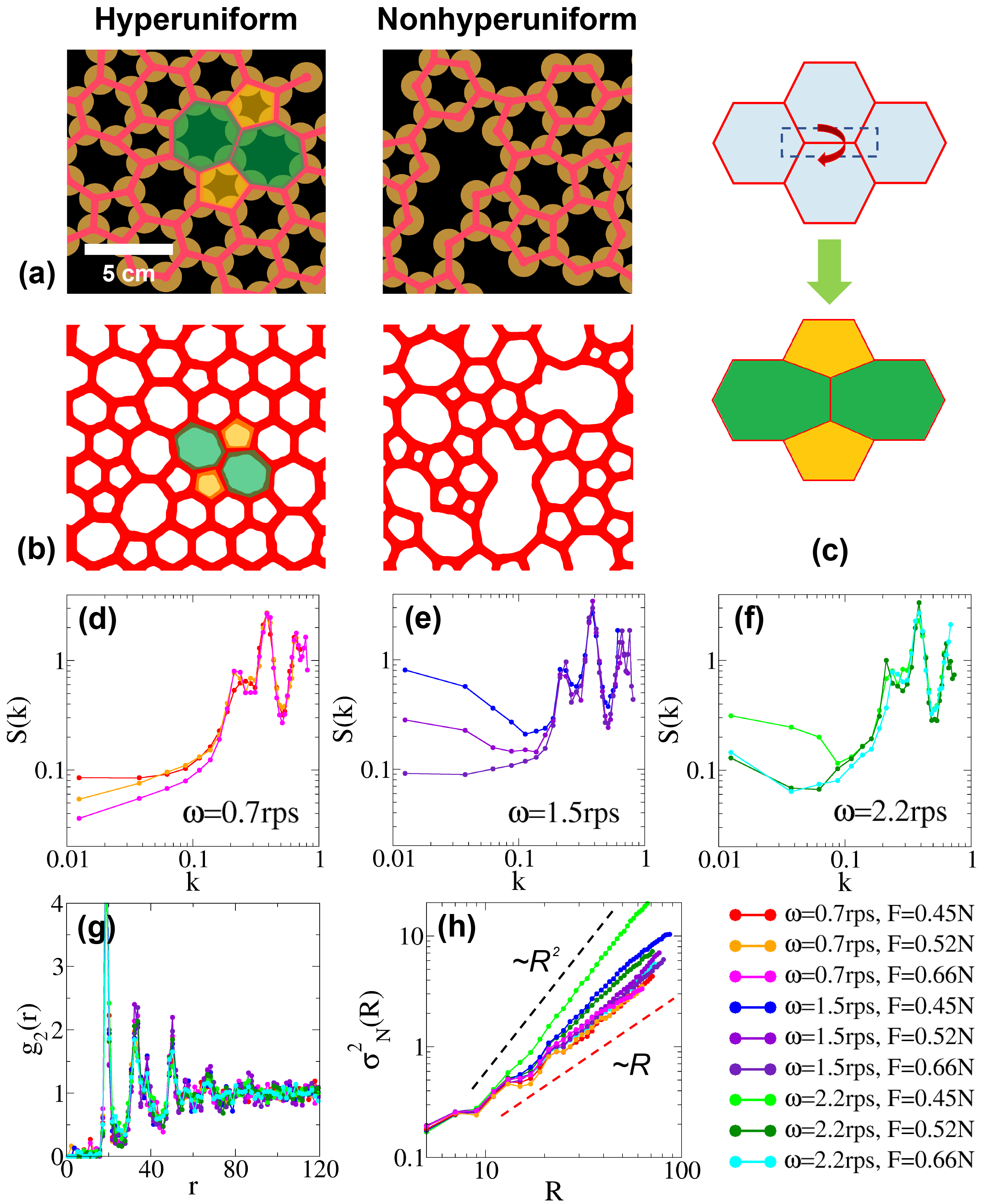} 
\end{array}$
\end{center}
\vspace{-0.7cm}
\caption{(a) Left panel: A hyperuniform network of Magbots (shown as disks) associated with $\omega = 0.7$ rps and $F = 0.52$ N (magnetic bonds are visualized). Right panel: A non-hyperuniform network associated with $\omega = 2.2$ rps and $F = 0.52$ N. (b) Simulated network obtained by introducing SW defects ($p = 0.10$) that is hyperuniform (left) and a network with both SW defects ($p = 0.10$) and vacancies ($c = 0.05$) that is non-hyperuniform (right). (c) Illustration of a Stone-Wales (SW) topological transformation. (d)-(f) respectively shows $S(k)$ of the self-assembled networks at $\omega = 0.7$, 1.5 and 2.2 rps for varying $F$. (g) and (h) respectively shows $g_2(r)$ and $\sigma_N^2(R)$ of the networks for varying $\omega$ and $F$.} \label{fig_3}
\end{figure}



\subsection{Emergence of active hyperuniform networks}

We obtain statistics of the point configurations derived from the network nodes (i.e., the Magbot positions), including the static structure factor $S(k)$, the number variance $\sigma_N^2(R)$, and the pair-correlation function $g_2(r)$. Fig. \ref{fig_3}a shows a hyperuniform network ($\omega = 0.7$ rps and $F = 0.52$ N) and a non-hyperuniform network ($\omega = 2.2$ rps and $F = 0.52$ N). The hyperuniform network mainly consists of well-defined 6-fold, 5-fold, and 7-fold rings of bonded Magbots, while the non-hyperuniform network contains distorted, broken rings and large voids (leading to large density fluctuations). 

A closer inspection reveals these hyperuniform networks contain well-defined Stone-Wales (SW) defects \cite{St86}. As illustrated in Fig. \ref{fig_3}c, a SW defect is generated by a topological transformation that converts 4 neighboring hexagons into 2 pentagons and 2 heptagons via a 90 degree bond rotation, leading to 5-fold and 7-fold rings of Magbots that are stabilized by the relatively stronger magnetic force $F$. These SW defects are commonly seen in atomic 2D materials, including amorphous silica \cite{Zh20} and graphene \cite{Ch21}, which are shown to preserve the hyperuniformity. To further illustrate this point, we numerically generated disordered network structures by continuously introducing SW defects in an initially perfect honeycomb network, following Ref. \cite{Ch21} (see SI Sec. 3). The amount of the SW defects is quantified via the defect contraction $p$, i.e., the number of flipped bonds over the total number of bonds. Fig. \ref{fig_3}b shows a simulated network containing SW defects with $p = 0.08$ that is hyperuniform and SW transformed network doped with randomly distributed vacancies, which destroy hyperuniformity \cite{chen2021topological}. In SI Sec. 3, we provide detailed statistics of these simulated networks and show they can structurally model the self-organized networks of the Magbots.   





Figures \ref{fig_3}d-f show $S(k)$ for the 9 sets of control parameters, grouped according to $\omega$. To quantify the degree of hyperuniformity, we employ the \textit{hyperuniformity index} \cite{To18a}, i.e., $H = \lim_{k\rightarrow 0}S(k)/S(k_{p})$, where $k_p$ is the wavevector associated the first and highest peak of $S(k)$. For a given value of $\omega$, increasing $F$ leads to a stronger suppression of $S(k)$ at $k\rightarrow 0$. In the case of $\omega = 0.7$ rps, the networks with $F = 0.52$ N and 0.66 N possess $H < 10^{-3}$ and a hyperuniformity exponent (obtained by fitting the small-$k$ data of $S(k)$) $\alpha \approx 0.32$ and 0.48 respectively, indicating these networks are hyperuniform to a high degree. The network with $F = 0.45$ N possesses $H < 10^{-2}$ and $\alpha \approx 0.14$. For the two larger $\omega$, except for the network with $\omega = 1.5$ rps and $F = 0.66$ N that is nearly hyperuniform (possessing $H \sim 10^{-2}$), all of the remaining active networks are non-hyperuniform, with $H \sim 1$. 



Figure \ref{fig_3}g shows $g_2(r)$, characterizing the short-range correlations. All systems possess a very strong first peak, corresponding to the three-coordinated local structure due to magnetic binding. The networks with larger $\omega$ have a faster decay and much smaller peaks at larger distances, indicating the degradation of the ``ring'' structures and emergence of large voids. Figure \ref{fig_3}h shows $\sigma_N^2(R)$ of the 9 active networks. The large-$R$ scaling behaviors are bounded between $\sigma_N^2(R) \sim R$ and $\sim R^2$. The hyperuniform systems associated with $\omega = 0.7$ rps possess the scaling $\sigma_N^2(R) \sim R^\beta$, respectively with $\beta \approx 1.51$ (for $F = 0.66$ N), $\beta \approx 1.64$ (for $F = 0.52$ N) and $\beta \approx 1.88$ (for $F = 0.45$ N), which are consistent with small-$k$ scaling of $S(k)$. The non-hyperuniform systems possess $\beta \approx 2$.

\subsection{Origin and robustness of hyperuniformity}

How does hyperuniformity emerge in the self-organized networks? As shown above, the key structural features leading to DHU are the SW defects. In the Appendix section, we show these SW defects are stabilized by the magnetic binding forces $F$ between the Magbots and destroyed by the local twist due to the active rotation as the speed $\omega$ increases (see Fig. 5a-b). Specifically, we analyze the self-organizing processes associated with distinct $F$-$\omega$ regimes, by quantifying the bond changing rate $\Gamma$ and the number of bonds $n_b$ per Magbot, which respectively characterizes the dynamics and structural evolution of the system (see Fig. 5d-e). These analyses reveal three distinct organizational behaviors: nonhyperuniform glass-like state, nonhyperuniform liquid-like state, and hyperuniform state, corresponding to binding-dominant, rotation-dominant, and balanced regimes (see Fig. 5c and Appendix for details).      


To further elucidate the role of interactions between the Magbots, we develop an active-particle model for the Magbots, which follow the underdamped Langevin equation:
\begin{equation}\label{eq3}
\begin{aligned}
m\frac{d\mathbf{v}}{dt}&= \mathbf{F}_a + \mathbf{F}_I + \mathbf{F}_w - \gamma_t \mathbf{v},\\
I\frac{d\omega}{dt}&= T_a + T_m + T_f - \gamma_r \omega,
\end{aligned}
\end{equation}
where $m$ and $I$ are mass and inertia, ${\bf v}$ and $\omega$ are translational and angular velocity, ${\bf F}_a$ and $T_a$ are the active force and torque, ${\bf F}_I$ is the total interaction force including core repulsion, magnetic interaction, and friction between the Magbots, ${\bf F}_w$ is due to the boundary confinement, $T_m$ and $T_f$ are torques due to magnetic and friction forces, $\gamma_t$ and $\gamma_r$ are translational and rotational friction coefficients, respectively (see SI Sec. 5-6 for details). 

\begin{figure}[ht!]
\begin{center}
$\begin{array}{c}\\
\includegraphics[width=0.45\textwidth]{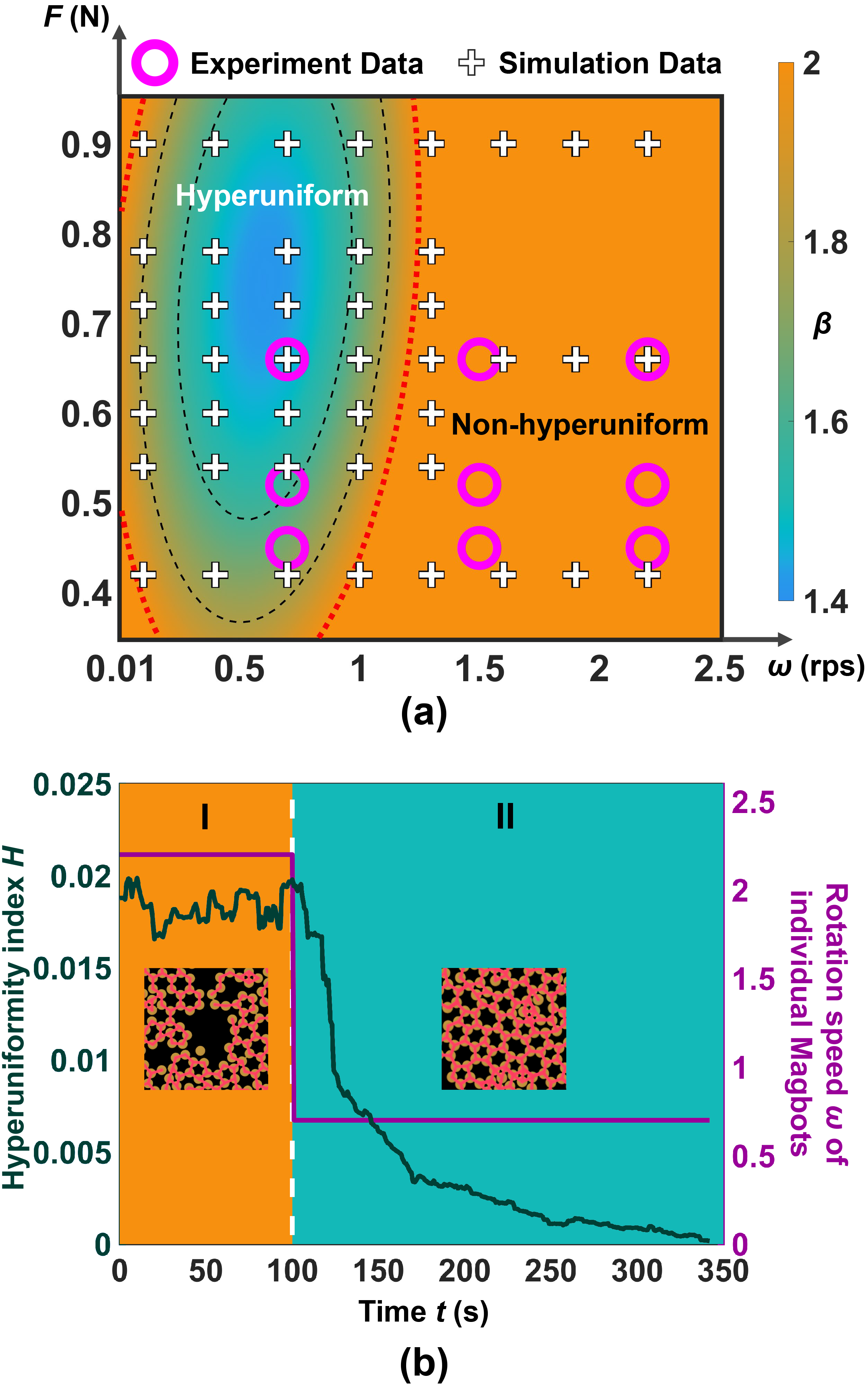} 
\end{array}$
\end{center}
\vspace{-0.8cm}
\caption{(a) A hyperuniform phase diagram of Magbots on the scaling exponent $\beta$ for $\sigma^2(R)$ informed by our experiments and simulations, where hyperuniform and nonhyperuniform phases are separated by the contour $\beta(F, \omega) = 2$ (red dashed line). (b) Transition of the disordered non-hyperuniform to the hyperuniform network structure (insets), quantified by the hyperuniformity index $H$ in response to the varying $\omega$.} \label{fig_4}
\end{figure}

This model allows us to systematically investigate the mechanism for the DHU networks, involving the competition between magnetic binding and local twist, beyond the values explored in the experiments. We select additional 44 pairs of $F$ and $\omega$ values complementary to the experimental data and numerically investigate the self-organization of these systems (see SI Sec. 7 for details). Based on the scaling exponent $\beta$ for $\sigma^2_N(R)$, which is well-defined in both hyperuniform and nonhyperuniform cases, we employ the piecewise cubic interpolation and quadratic polynomial surface fitting to obtain a continuous function $\beta(F, \omega)$, see SI Sec. 7. This allows us to construct a ``hyperuniformity phase diagram'' as shown in Fig. \ref{fig_4}a, where hyperuniform and nonhyperuniform phases are separated by the contour $\beta(F, \omega) = 2$. The phase diagram reveals that a variety of Class-III DHU networks possessing distinct structural characteristics (i.e., different $\beta \in (1.4, 2)$) robustly emerge in the upper left region corresponding to the balanced regime between magnetic binding and activation. These results also indicate one can controllably achieve distinct stable DHU structures via tuning $F$ and $\omega$.

Last but not least, we demonstrate the robustness of the hyperuniform state: We prepare a non-hyperuniform initial state with $F = 0.52$ N, which is an unflavored yet meta-stable configuration under $\omega = 0.7$ rps, i.e., an analog of the kinetically trapped non-equilibrium state. We then increase $\omega$ to 2.2 rps to kinetically activate the system out of the trap (e.g., breaking the bonds between undesirable neighbors), and then decrease $\omega$ to 0.7 rps to allow the Magbots to reorganize and form stable magnetic binding between the neighbors, which leads to a hyperuniform network. Fig. \ref{fig_4}b shows the evolution of hyperuniformity index $H$ in response to the varying $\omega$, which clearly reveals the reorganization of the Magbots to form hyperuniform structures.

\section{Conclusions}

Using a robotic system, we elucidated a new mechanism for achieving DHU states via the self-organization of active systems, involving the competition of reversible binding and symmetry breaking activation. It is also interesting to explore mixtures of different Magbot species, e.g., with different chirality or magnetic forces. The resulting distinct network structures would give rise to a wide spectrum of elastic wave propagation characteristics for phononic applications.




The authors thank the anonymous referees for their helpful comments and valuable suggestions. J. W., H. C., G. W. and L. L. were supported by NSFC (Grant No.T2350007, No. 12404239, No. 12174041), seed grants from the Wenzhou Institute (WIUCASQD2021002), and CPSF (Grant No. 2022M723118). G. C. was supported by Wenzhou Key Laboratory of Biophysics, Wenzhou Institute (Grant No. WIUCASSWWL22002). Y. J. did not have support on this work.

\appendix 

\section{Mechanisms for Stable DHU Networks}







\begin{figure}[ht!]
\begin{center}
$\begin{array}{c}
\includegraphics[width=0.45\textwidth]{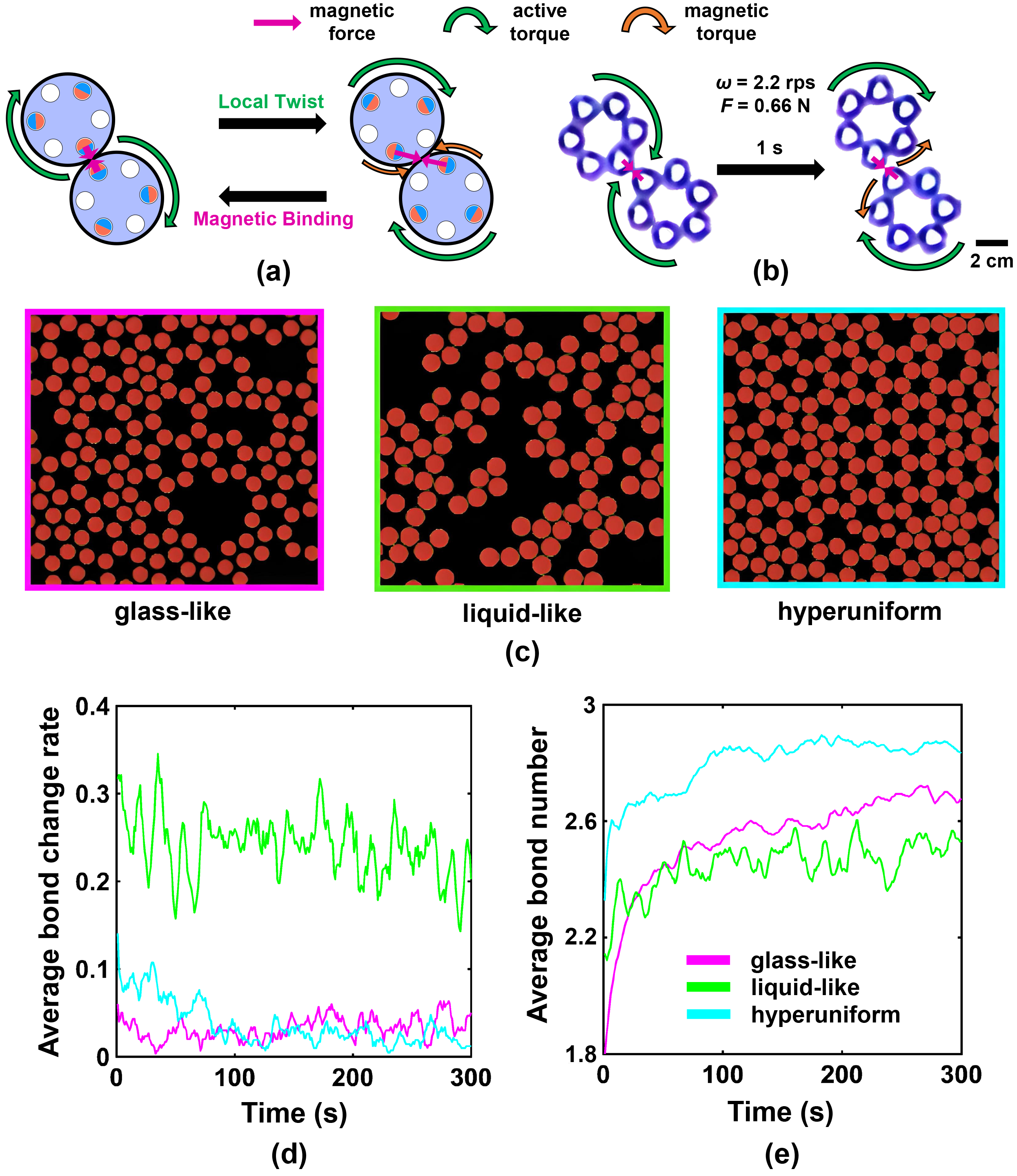} 
\end{array}$
\end{center}
\vspace{-0.7cm}
\caption{(a) Schematic illustration of the competition between the magnetic binding and local twist due to active rotation of Magbots. (b) Breaking of a spinner cluster due to the strong local twist at high rotation speed. (c) Representative configurations of glass-like, liquid-like and hyperuniform networks generated in simulations with $N \sim 1000$ Magbots. (d) Average bond change rate for different ``competition regimes'' between the magnetic binding forces and active forces. (e) Average bond number per Magbot as a function of time for different ``competition regimes''.} \label{fig_5}
\end{figure}

In this section, we describe the mechanisms to achieve the stable DHU self-organizations. As shown in Fig. \ref{fig_5}a, the active rotation of Magbots generates local twist that distorts the alignment of magnetic binding sites. This misalignment then leads to a magnetic torque, which competes with the active torque to restore the alignment. In SI Sec. 6, our simulations and analysis show that among other local interactions including friction and repulsive contact force, the magnetic binding and local twist due to active rotation are the two dominant factors determining the self-organization behaviors. At low rotation speed $\omega$, individual Magbots bind together by the magnetic force $F$ to form rotating clusters. As $\omega$ increases, the twist effect becomes stronger, which weakens and eventually destroys the ``bonds'' between the Magbots (see Fig. \ref{fig_5}b). Faster rotation also significantly increases the impact forces during the collisions between the clusters/individual Magbots, frustrating the ordering of the system.



In the binding-dominant regime (e.g., strong $F$, low $\omega$), the system tends to form a highly disordered, glass-like stable aggregate of Magbots containing holes and defects, which is non-hyperuniform (Fig. \ref{fig_5}c, left panel). In the rotation-dominant regime (e.g., weak $F$, high $\omega$), the system is highly dynamic, with clusters constantly forming and breaking, mimicking a liquid-like state and is non-hyperuniform (Fig. \ref{fig_5}c, middle panel). In a regime where magnetic binding (magnetic torque) and active twist (active torque) achieve a delicate balance, the weak bonds are eliminated and new stronger bonds are formed as the Magbots locally reorganize, a stable hyperuniform structure emerges (Fig. \ref{fig_5}c, right panel).

We compute the time-dependent average ``bond change rate'' $\Gamma(t)$ for the distinct regimes from experimental data, defined as number of bond breaking/forming per unit time, see Fig. \ref{fig_5}d. For the glass-like state ($F$-dominant, far left region of the phase diagram in Fig. \ref{fig_4}a), $\Gamma$ remains to be low during the evolution. For the liquid-like state ($\omega$-dominant, lower right region of the phase diagram), $\Gamma$ remains to be large during the evolution, indicating the system is highly dynamic. For the ``balanced'' regime, $\Gamma$ is initially large and gradually decays to a very small value, indicating the re-organization of the structure through weak-bond breaking and new-bond formation as discussed above. 

Fig. \ref{fig_5}e quantifies the structural evolution via the number of bonds per Magbots $n_b(t)$ from the experimental data. For DHU networks, $n_b = 3$ in the infinite-system limit, and is slightly smaller than 3 in finite systems (due to boundary effects). On the other hand, voids and vacancies result in $n_b$ significantly smaller than 3. Fig. \ref{fig_5}e shows that $n_b(t)$ for the DHU network initially increases (due to re-organizations that stable SW defects) and asymptotes at a value slightly smaller than 3. For the other two cases, $n_b(t)$ remains much smaller than 3, due to large voids or vacancies. Together with the structural analysis presented in the main text, these results provide quantitative evidence for the aforementioned competition mechanisms leading to the rich organizing behaviors of the Magbot system.


\bibliography{network}

\end{document}